# Point-contact Andreev reflection spectroscopy of heavy-fermion-metal/superconductor junctions

W. K. Park,[a,*] H. Stalzer,[a] J. L. Sarrao,[b] J. D. Thompson,[b] L. D. Pham,[c] Z. Fisk,[d] J. Frederick,[e] P. C. Canfield,[e] L. H. Greene[a]

[a]*Department of Physics, University of Illinois at Urbana-Champaign, Urbana, Illinois 61801, USA*

[b]*Los Alamos National Laboratory, Los Alamos, New Mexico 87545, USA*

[c]*Department of Physics, University of California, Davis, California 95616, USA*

[d]*Department of Physics and Astronomy, University of California, Irvine, California 92697, USA*

[e]*Ames Laboratory and Department of Physics and Astronomy, Iowa State University, Ames, Iowa 50011, USA*

**Abstract**

Our previous point-contact Andreev reflection studies of the heavy-fermion superconductor $CeCoIn_5$ using Au tips have shown two clear features: reduced Andreev signal and asymmetric background conductance. To explore their physical origins, we have extended our measurements to point-contact junctions between single crystalline heavy-fermion metals and superconducting Nb tips. Differential conductance spectra are taken on junctions with three heavy-fermion metals, $CeCoIn_5$, $CeRhIn_5$, and $YbAl_3$, each with different electron mass. In contrast with $Au/CeCoIn_5$ junctions, Andreev signal is not reduced and no dependence on effective mass is observed. A possible explanation based on a two-fluid picture for heavy fermions is proposed.

According to the Blonder-Tinkham-Klapwijk theory [1], Andreev reflection (AR) [2] process is prohibited if the two electrodes have highly disparate Fermi velocities, as in normal-metal/heavy-fermion superconductor junctions. However, a conductance enhancement due to AR has been frequently observed in point-contact junctions with heavy-fermion superconductors [3,4], including our results on $CeCoIn_5$ [5], albeit reduced [3-5]. Deutscher and Nozières addressed this discrepancy by proposing that relevant boundary conditions are without mass enhancement factors [6]. Although this theory provides an explanation for why AR is observable in heavy-fermion superconductors, the role played by the large effective electron mass ($m^*$) in the AR process is still not understood [5].

In order to address this question, we have carried out conductance measurements on junctions with three heavy-fermion metals (HFN), each with different $m^*$ value: $CeCoIn_5$ ($T_c$ = 2.3 K, $m^* \sim 83 m_0$) [7], $CeRhIn_5$ ($5$-$9 m_0$, below $T_N$ = 3.8 K) [8], and $YbAl_3$ ($15$-$30 m_0$) [9]. Point-contact junctions are made by bringing electrochemically prepared superconducting Nb tips onto the (001) surfaces of solution-grown [10] heavy-fermion single crystals [11]. Differential conductance spectra are taken by standard lock-in techniques. Here, positive voltage means that the HFN electrode is biased positively.

Normalized conductance spectra for Au/Nb and $CeCoIn_5$/Nb junctions are displayed in Fig. 1. They appear similar qualitatively with double peak structures due to the superconducting energy gap of Nb. The peak amplitude in the $CeCoIn_5$/Nb junction appears smaller than in the Au/Nb junction, which may be due to the larger quasiparticle lifetime smearing effect in $CeCoIn_5$ as evidenced by the more rounded peak shape. Note the conductance asymmetry in the $CeCoIn_5$/Nb junctions, similar to that observed in $Au/CeCoIn_5$ junctions [5].

In Fig. 2, normalized conductance data for three HFN/Nb junctions are plotted together. The observed conductance asymmetry seems to be a common behavior in heavy-fermion junctions, implying that it may be due to an energy-dependence of the density of states. As for the AR signal, no clear dependence on the effective mass is observed in these three junctions. An experimental observation not shown here is that the AR conductance in $CeRhIn_5$/Nb

---

junctions does not exhibit any signatures for the antiferromagnetic transition with which the electron mass is supposed to show a non-monotonic behavior. There is no strong correlation between the observed AR signal and the effective electron mass and the AR signal in HFN/Nb junctions is not reduced, in contrast with Au/CeCoIn$_5$ junctions [5].

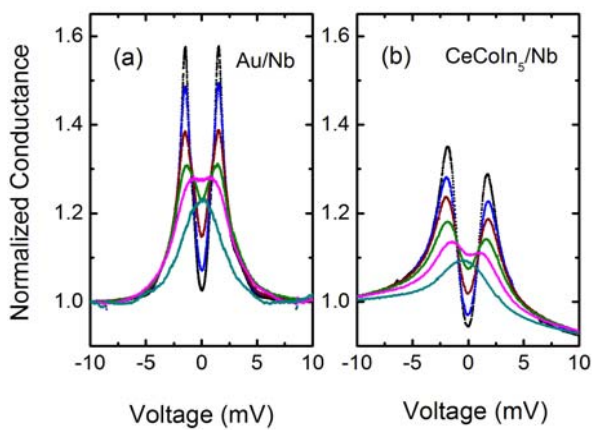

Fig. 1: (color online). Normalized conductance spectra of (a) Au/Nb and (b) CeCoIn$_5$/Nb junctions. (a) Temperatures are 2.2, 3.1, 4.4, 6.0, 7.5, and 9.1 K from the top in the peak position. (b) Temperatures are 1.9, 3.8, 5.3, 6.8, 8.0, and 9.0 K from the top in the peak position.

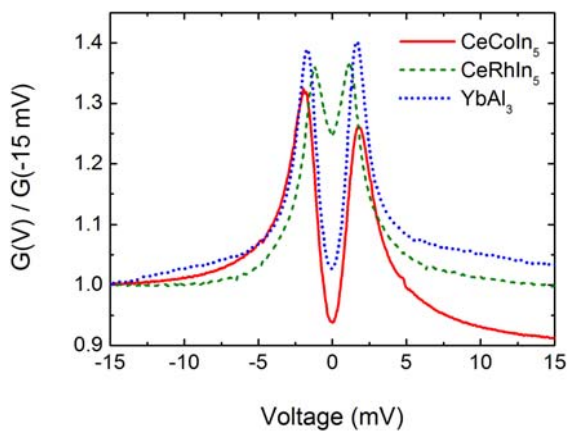

Fig. 2: (color online). Normalized conductance spectra of heavy-fermion-metal/Nb junctions. The solid curve is for CeCoIn$_5$ (2.5 K), dashed one for CeRhIn$_5$ (1.83 K), and dotted one for YbAl$_3$ (2.6 K).

To understand these contrasting behaviors, we consider the two-fluid picture [12] proposed for the emergent heavy fermion liquid in a Kondo lattice system. We also note the report on the existence of unpaired light electrons below $T_c$ of CeCoIn$_5$ [13]. Then, a possible explanation for the reduced AR signal is the non-participation of the uncondensed light electrons in the AR process. In contrast, both heavy and light electrons are expected to participate in the AR process in the CeCoIn$_5$/Nb junctions, giving a usual AR signal.

This work was supported by the U.S. DoE, Award DEFG02-91ER45439, through the FSMRL and the CMM at UIUC, by NSF-DMR-0503360 at UCD & UCI, and performed at LANL under auspices of the U.S. DoE, Office of Science. H.S. acknowledges the support from the Deutsche Forschungsgemeinschaft.